\documentstyle[12pt]{article}
\newcommand{\eq}{\begin{equation}}
\newcommand{\en}{\end{equation}}
\newcommand{\eqn}{\begin{eqnarray}}
\newcommand{\enn}{\end{eqnarray}}

\begin{document}
\begin{titlepage}
\begin{flushright}
THEP-93-3 \\
Feb. 1993
\end{flushright}
\vspace{2cm}
\begin{center}
{\LARGE
Hermitian Jordan Triple Systems, \\
the Standard Model plus Gravity, and  \\
$\alpha_{E} = 1/137.03608$ } \\

\vspace{2cm}
{\large Frank D. (Tony) Smith, Jr.} \\
Department of Physics \\
Georgia Institute of Technology \\
Atlanta, Georgia 30332 \\
\vspace{1cm}
{\bf Abstract}
\end{center}
A physical interpretation is given for some
Hermitian Jordan triple systems (HJTS) that were
recently discussed by G\"{u}naydin \cite{MG2}.

The quadratic Jordan algebras derived from HJTS
provide a formulation of quantum mechanics
(G\"{u}naydin\cite{MG1}) that is a natural framework
within which exceptional structures are identified
with physically realistic structures of a
quantum field theory that includes both
the standard model and MacDowell-Mansouri
\cite{MM} gravity.

The structures allow the calculation of the relative
strengths of the four forces, including
$\alpha_{E} = 1/137.03608$.

\vspace{1cm}
\normalsize
\footnoterule
\noindent
{\footnotesize \copyright 1993 Frank D. (Tony) Smith, Jr.,
341 Blanton Road, Atlanta, Georgia 30342 USA \\
P. O. Box for snail-mail:  P. O. Box 430, Cartersville,
Georgia 30120 USA \\
e-mail: gt0109e@prism.gatech.edu
and fsmith@pinet.aip.org}
\end{titlepage}

\newpage

\setcounter{footnote}{0}
\section{Introduction}
\setcounter{equation}{0}

\subsection{Overview}

Since the purpose of this paper is to give a physical
quantum field theoretical interpretation of the
algebraic structures (Jordan algebras, Lie algebras,
symmetric spaces, bounded homogeneous domains, etc.)
the field over which the structures are defined is
the field $\bf C$ of the complex numbers.

In particular, the initial fundamental Lie algebras
will be complexifications $G^{\bf C}$
of real Lie algebras $G$.

Real structures, such as spacetime, will emerge
in the physical interpretation by such mechanisms
as by taking the Silov boundary of a bounded complex
homogeneous domain.

\vspace{12pt}

The remainder of this first section of this paper,
$Introduction$, will be devoted to an overview of the
algebraic and geometric structures to be used.  Also,
the work of G\"{u}naydin\cite{MG1} comparing the
quantum mechanical formalisms of quadratic and
bilinear Jordan algebras with the conventional
matrix-operator on Hilbert space formalism is used
to justify the quadratic Jordan algebra approach
taken in this paper.

\vspace{12pt}

The second section, $Physical \: Interpretation$,
will deal with the physical interpretation of some
specific structures that are taken to be fundamental.

\vspace{12pt}

In the third section, $Calculation \: of \: Observable
\: Quantities$, explicit calculations will be made
of physically measurable quantities, such as the
electromagnetic fine structure constant
$\alpha_{E} = 1/137.03608$.

\vspace{12pt}

Since the structures of this paper are defined over the
field of complex numbers, the difference between
Minkowski signature (d-1, 1) and Euclidean signature (d, 0)
is not immediately apparent, because Wick rotation
of complex spaces can transform back and forth
between Minkowski and Euclidean signature.

However, after working with the structures
(using such processes as taking Silov boundaries
of bounded complex homogeneous domains) to get
some  physical structures such as spacetime,
structures defined over the real numbers will emerge.

For example, the Lie algebra of a physical
gauge group will be the real Lie algebra $G$
underlying its complexification $G^{\bf C}$.

In such cases, signature of structures can be defined.
In this paper, calculations of physically observable
quantities use the ratios of volumes of geometric spaces.
Therefore, the spaces should be represented in forms
in which their volumes are finite.  Instead of non-compact
forms arising from vector spaces with Minkowski signature
(d-1, 1), compact forms arising from vector
spaces with Euclidean signature (d, 0) are used.

It is assumed in this paper that any physics requiring
spaces with Minkowski signature can be obtained
from the spaces with Euclidean signature by Wick
rotation of complex spaces (which are taken to be
fundamental).

\subsection{Jordan Triple Systems}

A Jordan triple system is defined by Satake \cite{STK}
as a finite-dimensional vector space $V$ of
dimension d (usually complex in this paper) with a
trilinear triple product map $$ \{  \} : V \times V
\times V \rightarrow V$$
such that for all $a,b,x,y,z \epsilon V$:

$$\{ x,y,z \} = \{ z,y,x \}$$
and
$$\{ a,b, \{ x,y,z \} \} =  \{ \{ a,b,x \} ,y,z \}
- \{ x, \{ b,a,y \} ,z \} + \{ x,y, \{ a,b,z \} \}$$

\subsection{Bilinear and Quadratic Jordan Algebras}

When there is a unit element $e$, the triple product can be
used to define two other different products:

a bilinear product $x \circ y =\{ x,y,e \} = \{ x,e,y \} = \{ e,x,y, \}$;
and

a quadratic product $U_{x} y = \{ x,y,x \}$.

\vspace{12pt}

In terms of the ordinary matrix  product $xy$ for
vectors $x$ and $y$ in the vector space $V$, the
products are (up to factors of $2$ or $1 \over 2$)
given by McCrimmon \cite{McC} as:

\vspace{12pt}

Triple product:  $$\{ x,y,z \} = xyz + zyx$$

\vspace{12pt}

Bilinear product:  $$x \circ y = {1 \over 2}(xy+yx)$$

\vspace{12pt}

Quadratic product:  $$U_{x} y = xyx$$

As McCrimmon \cite{McC} states, the bilinear and quadratic
Jordan algebras are categorically equivalent when there
exists a scalar $1 \over 2$.

\subsection{Quadratic Jordan Algebra Quantum Mechanics}

G\"{u}naydin \cite{MG1} discusses the use of either the
bilinear Jordan product or the quadratic Jordan
product to formulate quantum mechanics in an
alternative way to the usual Hilbert space formulation.

The following chart taken from G\"{u}naydin \cite{MG1}
compares various quantities in the three formulations:

\[
\begin{array}{|c|c|c|}
\hline
Hilbert \:Space & Bilinear \:Jordan &  Quadratic \:Jordan \\
\hline
&& \\
 | \alpha \rangle & | \alpha \rangle \langle \alpha | = P_{\alpha} &
P_{\alpha}  \\
&& \\
H  | \alpha \rangle  & H \circ P_{\alpha} &
\Pi_{H} P_{\alpha} =  \{H P_{\alpha} H \} \\
 &&  \\
\langle \alpha | H  | \beta \rangle & ? & ?  \\
&& \\
\langle \alpha | H  | \alpha \rangle & Tr \: H \circ P_{\alpha} &
Tr \: \Pi_{P_{\alpha}} H =  Tr \: \{ P_{\alpha} H P_{\alpha} \}    \\
&& \\
| \langle \alpha | H  | \beta \rangle |^{2} &
Tr \: P_{\alpha} \circ \{ H P_{\beta} H \}  &
Tr \: \Pi_{P_{\alpha}} \Pi_{H} P_{\beta}   \\
& = Tr \: \{ H P_{\alpha} H \}  \circ  P_{\beta} &
= Tr \: \Pi_{P_{\beta}} \Pi_{H} P_{\alpha} \\
&& \\
{[}H_{1} , H_{2}{]}& ? &
4(\Pi_{H_{1}} \circ \Pi_{H_{2}} -
\Pi_{H_{1} \circ H_{2}})        \\
\hline
\end{array}
\]

Neither the bilinear Jordan formulation nor the quadratic
Jordan formulation has a natural equivalent to the
Hilbert space formulation of the transition matrix element
$\langle \alpha | H  | \beta \rangle$.

Since $\langle \alpha | H  | \beta \rangle$ is not measurable,
the two Jordan formulations are closer to physical reality
than the Hilbert space formulation.

\vspace{12pt}

In the bilinear Jordan formulation, there is no natural direct
analog of the commutator of two observables $[H_{1} , H_{2}]$.
According to G\"{u}naydin \cite{MG1}, that problem is usually
managed by requiring the associator of the two observables
with all elements of the bilinear Jordan algebra to vanish.

\vspace{12pt}

As G\"{u}naydin \cite{MG1} states, there is an analog of the
commutator in the quadratic Jordan formulation:
$$4(\Pi_{H_{1}} \circ \Pi_{H_{2}} - \Pi_{H_{1} \circ H_{2}})$$
Therefore, in the quadratic Jordan formulation  two observables
are compatible when
$\Pi_{H_{1}} \circ \Pi_{H_{2}} = \Pi_{H_{1} \circ H_{2}}$.

\vspace{12pt}

In this paper, the quadratic Jordan formulation of quantum
mechanics is used for the physical interpretation of Jordan
triple systems.

\subsection{Jordan Pairs:  Conformal and Lorentz Algebras}

McCrimmon \cite{McC} says that Jordan triple systems
with vector space $V$ are in 1-1 correspondence with
Jordan pairs $(V_{+} , V_{-})$  with involution such
that $V_{+} = V_{-} = V$.

\vspace{12pt}

If $V$ is d-dimensional then the Jordan pairs $(V_{+} , V_{-})$
with involution can be used to construct the conformal Lie
algebra of the vector space $V$, following the approaches
discussed by G\"{u}naydin \cite{MG2} and by McCrimmon
\cite{McC}.
(Here, anticipating the need for compact Euclidean spaces
in calculations, the d-dimensional vector space $V$ is
represented by  $S^{d}$, the d-dimensional sphere.):

\vspace{12pt}

the translations of $V = S^{d}$ are $V_{+} = S^{d}_{+}$;

\vspace{12pt}

the special conformal transformations of $V = S^{d}$ are
$V_{-} = S^{d}_{-}$;

\vspace{12pt}

the Lorentz transformations and dilatations of $V = S^{d}$ are the

Lie bracket $[V_{+} , V_{-}] = [S^{d}_{+}, S^{d}_{-}]$; and

\vspace{12pt}

the full conformal Lie algebra is the 3-graded Lie algebra
$$V_{-} \oplus [V_{+} , V_{-}] \oplus V_{+} =$$
$$=S^{d}_{-} \oplus [S^{d}_{+}, S^{d}_{-}] \oplus S^{d}_{+}$$
A conventional example (using real vector spaces and
uncomplexified real Lie algebras) is the conformal Lie
algebra $Spin(6)$ where the vector space $V$ is
4-dimensional with Euclidean signature (4, 0):
$$Spin(6) =$$
$$= S^{4}_{-} \oplus (Spin(4) \otimes U(1))
\oplus S^{4}_{+}$$

\vspace{12pt}

G\"{u}naydin \cite{MG2} describes the conformal
Lie algebras and Lorentz Lie algebras of simple
Jordan algebras.

A similar description (substituting some compact
forms for non-compact ones) is as follows, denoting
by  $J^{\bf C}_{n}(A)$ and by $\Gamma(d)$ the Jordan
algebra of $n\times n$  Hermitian matrices over the
division algebra $A$ and the Jordan algebra of Dirac
gamma matrices in $d$ complex dimensions, and
denoting the four division algebras
(real numbers, complex numbers, quaternions,
and octonions)
by the symbols $ {\bf R} $, ${\bf C}$, ${\bf H}$,
and ${\bf O}$.

This chart shows:

the Jordan (grade -1) algebra;

the Lorentz (grade 0) Lie algebra plus
the dilatation algebra $U^{\bf C}(1)$;

the other member of the Jordan pair, the
Jordan (grade +1) algebra; and

the conformal Lie algebra that is the
sum of all three grades.

\[
\begin{array}{|c|c|c||c|}
\hline
&& \\
Jordan \:(-1) & Lorentz \oplus Dilatation \:(0) & Jordan \:(+1)
& Conformal \\
\hline
&&& \\
J^{\bf C}_{n}(R) & SU^{\bf C}(n) \oplus U^{\bf C}(1)
& J^{\bf C}_{n}(R) & Sp^{\bf C}(2n)\\
&&& \\
J^{\bf C}_{n}(C)  & SU^{\bf C}(n) \times SU^{\bf C}(n)
\oplus U^{\bf C}(1) &J^{\bf C}_{n}(C)  &
SU^{\bf C}(2n) \\
&&& \\
J^{\bf C}_{n}(H)  & SU^{\bf C}(2n) \oplus U^{\bf C}(1)
& J^{\bf C}_{n}(H)  & SO^{\bf C}(4n) \\
&&& \\
J^{\bf C}_{3}(O)  & E^{\bf C}_{6} \oplus U^{\bf C}(1)
& J^{\bf C}_{3}(O)  & E^{\bf C}_{7} \\
&&& \\
\Gamma^{\bf C}(n) & Spin^{\bf C}(n+1) \oplus U^{\bf C}(1)
& \Gamma(n) & Spin^{\bf C}(n+3) \\
\hline
\end{array}
\]

\subsection{Hermitian Jordan Triple Systems (HJTS)}

As McCrimmon \cite{McC} states, a Jordan triple
system is an Hermitian Jordan triple system (HJTS)
if the triple product $\{ x,y,z \}$ is $ {\bf{C}} -linear$
in x and z but ${\bf{C}} -antilinear$ in y, and the
bilinear form $ \langle x,y \rangle = Tr \:  V_{x,y}$,
where $V_{x,y}: z \rightarrow \{ x,y,z \}$, is a positive
definite Hermitian scalar product.

\vspace{12pt}

There exist  four infinite families
of HJTS and two exceptional ones.
They are given by G\"{u}naydin \cite{MG2} as:

\vspace{12pt}

$Type \: I_{p,q}$ generated by $p\times q$ complex matrices
$M_{p,q}({\bf C})$ with the triple product
\begin{equation}
(abc) = a b^{\dagger} c + c b^{\dagger} a
\end{equation}
where $\dagger$ represents the usual hermitian conjugation.

\vspace{12pt}

$Type \: II_{n}$
 generated by complex anti-symmetric
  $n \times n$ matrices $A_{n}({\bf C})$ with the triple product
\begin{equation}
(abc) = a b^{\dagger} c + c b^{\dagger} a
\end{equation}
where $\dagger$ represents the usual hermitian conjugation.

\vspace{12pt}

$Type \: III_{n}$ generated by complex $n \times n$
 symmetric matrices $S_{n}({\bf C})$ with the triple product
\begin{equation}
(abc) = a b^{\dagger} c + c b^{\dagger} a
\end{equation}
where $\dagger$ represents the usual hermitian conjugation.

\vspace{12pt}

$Type \: IV_{n}$ generated by Dirac gamma matrices
$\Gamma^{\bf C}(n)$ in n dimensions
with complex coefficients and the Jordan triple product
\eq
(abc)= a\cdot (\bar{b} \cdot c) + c \cdot (\bar{b} \cdot a)
-(a \cdot c) \cdot \bar{b}
\en
where the bar $\bar{\;}$ denotes complex conjugation.

\vspace{12pt}

$Type \: V$ generated by $1 \times 2 $ complexified
octonionic matrices $M^{\bf C}_{1,2}({\bf O})$ with the
triple product
\begin{equation}
(abc) = \{ (a \bar{b}^{\dagger}) c + (\bar{b} a^{\dagger}) c
- \bar{b} (a^{\dagger} c) \} + \{ a \leftrightarrow c \}
\end{equation}
where $\dagger$ denotes octonion conjugation times
transposition and the bar $\bar{\;}$ denotes complex conjugation.

\vspace{12pt}

$Type \: VI$ generated by the exceptional Jordan algebra
of $3\times 3$ hermitian complexified octonionic matrices
$J^{\bf C}_{3}({\bf O})$ with the triple product
\eq
(abc)= a\cdot (\bar{b} \cdot c) + c \cdot (\bar{b} \cdot a)
-(a \cdot c) \cdot \bar{b}
\en
where the bar $\bar{\;}$ denotes complex conjugation.

\vspace{12pt}

The simple HJTS, their Jordan algebras,
their Lorentz Lie algebras,
and their Conformal Lie algebras are
as follows
(see G\"{u}naydin \cite{MG2} and
McCrimmon \cite{McC}):

\[
\begin{array}{|c|c|c|c|} \hline
HJTS & Jordan & Lorentz & Conformal  \\ \hline
&&& \\
S_{n}({\bf C}) & J^{\bf C}_{n}(R) & SU^{\bf C}(n)
& Sp^{\bf C}(2n) \\
&&& \\
M_{p,q}({\bf C}) & (If \: p=q=n): J^{\bf C}_{n}(C)
& SU^{\bf C}(p) \times SU^{\bf C}(q) & SU^{\bf C}(p+q)  \\
&&&  \\
A_{n}({\bf C}) & J^{\bf C}_{n}(H) & SU^{\bf C}(n)
& SO^{\bf C}(2n) \\
&&&  \\
J_{3}^{{\bf C}}({\bf O})  & J^{\bf C}_{3}({\bf O})
& E^{\bf C}_{6}  & E^{\bf C}_{7}  \\
&&& \\
\Gamma^{\bf C}(n) & \Gamma^{\bf C}(n) & Spin^{\bf C}(n+1)
& Spin^{\bf C}(n+3)\\
&&& \\
M^{\bf C}_{1,2}({\bf O}) & M^{\bf C}_{1,2}({\bf O}) &
Spin^{\bf C}(10) & E^{\bf C}_{6} \\ \hline
\end{array}
\]

\subsection{HJTS are 1-1 with \protect\\Bounded
Homogeneous Circled Domains}

McCrimmon states in Theorem 6.7 of \cite{McC},
there is a 1-1 correspondence between HJTS and
bounded homogeneous circled domains.

\subsection{HJTS are 1-1 with Hermitian Symmetric Spaces}

There is a 1-1 correspondence among HJTS, bounded
homogeneous circled domains and Hermitian symmetric
spaces, provided that the Hermitian symmetric spaces
of compact type are considered to be in the same equivalence
class as their noncompact dual Hermitian symmetric spaces.
\cite{STK,McC,HLG}

\vspace{12pt}

For the purpose of physical interpretation in this paper,
an HJTS is considered to be identified by the 1-1
correspondences just mentioned with the corresponding
bounded homogeneous circled domain and with the
corresponding compact Hermitian symmetric space.

\subsection{Which HJTS should be taken as Fundamental?}

If all the above HJTS can be used as the basis for a
quantum theory, which one should be taken as the
fundamental structure for interpretation as a physically
realistic theory?

\vspace{12pt}

Dixon \cite{GD}, noting that the only good division
algebras are ${\bf R}$, ${\bf C}$, ${\bf H}$, and ${\bf O}$,
and that nature ought to use all the good things
that are available, has advocated using the 64(real)
dimensional space ${\bf R} \otimes {\bf C} \otimes
{\bf H} \otimes {\bf O}$ as the fundamental
structure.

Since I like that approach, I note that the closest
thing to ${\bf R} \otimes {\bf C} \otimes{\bf H}
\otimes {\bf O}$ that is an HJTS is the HJTS of
$Type \: VI$, $J_{3}^{{\bf C}}({\bf O})$, the
$3 \times 3$ Hermitian matrices of octonions over
the complex number field.

\vspace{12pt}

$J_{3}^{{\bf C}}({\bf O})$ is 54(real) dimensional, and
can be represented as the complexification of the
following 27(real) dimensional matrix, where
${\bf O}_{+}, {\bf O}_{-}, {\bf O}_{v} $ are octonion,
$a, b, c$ are real, and $\dagger$ denotes octonion
conjugation:

\[
\left(
\begin{array}{ccc}
a & {\bf O}_{+} &  {\bf O}_{v} \\
& & \\
{\bf O}_{+}^\dagger & b &  {\bf O}_{-} \\
& & \\
{\bf O}_{v}^\dagger & {\bf O}_{-}^\dagger & c
\end{array}
\right)
\]

For details about the 27(real) dimensional
Jordan algebra $J_{3}({\bf O})$ see, for example,
Adams \cite{JFA}.

\vspace{12pt}

If $J_{3}^{{\bf C}}({\bf O})$ were to be equal to
${\bf R} \otimes {\bf C} \otimes{\bf H}
\otimes {\bf O}$, the matrix structure would
have to be equivalent to quaternionic structure.
Of course, it is not, but it contains the imaginary
part of the quaternions in the sense that the
imaginary quaternions $\bf i,j,k$ are contained in
the following 3(real) dimensional rotation matrices:

\[
\begin{array}{ccc}
\left(
\begin{array}{ccc}
a & {\bf O}_{+} &  0 \\
& & \\
{\bf O}_{+}^\dagger & b &  0 \\
& & \\
0 & 0 & c
\end{array}
\right)
&
\left(
\begin{array}{ccc}
a & 0 &  0 \\
& & \\
0 & b &  {\bf O}_{-}  \\
& & \\
0 & {\bf O}_{-}^\dagger & c
\end{array}
\right)
&
\left(
\begin{array}{ccc}
a & 0 &   {\bf O}_{v} \\
& & \\
0 & b &  0 \\
& & \\
{\bf O}_{v}^\dagger & 0 & c
\end{array}
\right)
\end{array}
\]

Since the $Type \: VI$ HJTS $J_{3}^{\bf C}
({\bf O})$ carries the most of ${\bf R} \otimes
{\bf C} \otimes {\bf H} \otimes {\bf O}$ that is
consistent with Jordan algebra structure, the
physical interpretation in this paper starts with
it, and with its corresponding
$Type \: VI$ Hermitian symmetric space of
compact type:

the set of
${\bf{(C \otimes O)}}P^2$'s in ${\bf{(H \otimes O)}}P^2$;

that is, ${E_{7}} \over {E_{6}
\times U(1)}$ with 54(real) dimensions.
\cite{HLG,ABS}

\vspace{12pt}

Note that the $Type \: VI$ Hermitian symmetric space
$${{E_{7}} \over {E_{6} \times U(1)}} =
{J_{3}^{\bf C}({\bf O})}$$
has 54 real dimensions,

whereas the Jordan pair for the $Type \: VI$ HJTS is
$${{E^{\bf C}_{7}} \over {E^{\bf C}_{6} \times U^{\bf C}(1)}} =
{J_{3}^{\bf C}({\bf O}) \oplus J_{3}^{\bf C}({\bf O})}$$
which has 54 complex dimensions, or 108 real dimensions.

\vspace{12pt}

Details of the physical interpretation are
discussed in the next section.

\section{Physical Interpretation}

\subsection{HJTS of $Type \: VI$:
$J^{\bf C}_{3}({\bf O})$}

The full conformal Lie algebra corresponding
to the HJTS of $Type \: VI$ is
the 3-graded Lie algebra
$$E^{\bf C}_{7} =$$
$$= J^{\bf C}_{3}({\bf O}) \oplus
(E^{\bf C}_{6} \times U^{\bf C}(1))
\oplus J^{\bf C}_{3}({\bf O})$$

\vspace{12pt}

The translations and the special conformal
transformations are each represented by the
space $J^{\bf C}_{3}({\bf O})$ of the HJTS
of $Type \: VI$.

\subsubsection{Space of Operators =
$J^{\bf C}_{3}({\bf O})$}
The space of operators of the quadratic Jordan
algebra quantum theory is identified with
the 54(real) dimensional space
$J^{\bf C}_{3}({\bf O})$.

\subsubsection{Provisional Space of States =
$J^{\bf C}_{3}({\bf O})$}
The provisional candidate for the quantum state
space is $J^{\bf C}_{3}({\bf O})$ itself.

This choice takes advantage of one distinction between
the quadratic Jordan algebra approach and the Hilbert
space matrix operator approach:

in the quadratic Jordan algebra approach, states are
projection operators $P_{\alpha}$ and the action of
the operator on the state is the quadratic Jordan
product $\Pi_{H} P_{\alpha} =  \{H P_{\alpha} H \}$;

in the Hilbert space matrix operator approach,
the states $ | \alpha \rangle$ is a
$3 \times 1$ (octonionic) vector space and the
action of the operator $3 \times 3$ matrices on the
state is the matrix product $H  | \alpha \rangle$.

\vspace{12pt}

The provisional state space is therefore
$J_{3}^{\bf C}({\bf O})$, which can be represented
as the following 54(real) dimensional matrix, where
${\bf O}^{\bf C}_{+}, {\bf O}^{\bf C}_{-}, {\bf O}^{\bf C}_{v}$
 are complexified octonions, $a^{\bf C}, b^{\bf C},
c^{\bf C}$ are complexified real numbers, and
$\dagger$ denotes octonion conjugation.

(Note that here the complexification was done after
the 27(real) dimensional $J_{3}({\bf O})$ was formed by
using the $\dagger$ of octonion conjugation for
Hermitianness.  This is different from forming
$J_{3}({\bf (C \otimes O})$ by using both the $\dagger$
of octonion conjugation and the $\bar{\;}$ of complex
conjugation for Hermitianness.

Also note that the quadratic Jordan product in
$J_{3}^{\bf C}({\bf O})$, and the triple product from
which it is derived, do use the $\bar{\;}$ of complex
conjugation.)

\[
\left(
\begin{array}{ccc}
a^{\bf C} & {\bf O}^{\bf C}_{+} &  {\bf O}^{\bf C}_{v} \\
& & \\
{\bf O}^{{\bf C}{\dagger}}_{+} & b^{\bf C}
&  {\bf O}^{\bf C}_{-} \\
& & \\
{\bf O}^{{\bf C}{\dagger}}_{v} &
{\bf O}^{{\bf C}{\dagger}}_{-} & c^{\bf C}
\end{array}
\right)
\]

\vspace{12pt}

The provisional physical interpretation of the
elements of the $3 \times 3$ matrix
$J_{3}^{{\bf C}}({\bf O})$ is:

\vspace{12pt}

${\bf O}^{\bf C}_{+}$ represents the (first generation)
fermion particles.  If a basis (with complex scalars)
for ${\bf O}^{\bf C}_{+}$ is
$\{ 1,i,j,k,e,ie,je,ke \} $,
then the representation is

\[
\begin{array}{|c|c|} \hline
{\bf O}^{\bf C}_{+}  & Fermion \: Particle \\
basis \: element & \\ \hline
1 & e-neutrino  \\ \hline
i & red \: up \: quark \\ \hline
j & green \: up \: quark \\ \hline
k & blue \: up \: quark \\ \hline
e & electron \\ \hline
ie & red \: up \: quark \\ \hline
je & green \: up \: quark \\ \hline
ke & blue \: up \: quark \\ \hline
\end{array}
\]

\vspace{12pt}

${\bf O}^{\bf C}_{-}$ represents the (first generation)
fermion antiparticles in the same physically
realistic way as the fermion particles were
represented;

\vspace{12pt}

${\bf O}^{\bf C}_{v}$ represents an 8-dimensional
spacetime, which can be reduced to 4 dimensions
as described in \cite{SM1,SM2}; and

\vspace{12pt}

$a^{\bf C}, b^{\bf C}, c^{\bf C}$ are not given a
physical interpretation.

\vspace{12pt}

\subsubsection{Difficulties with Provisional
Space of States = $J^{\bf C}_{3}({\bf O})$}
The provisional space of states $J^{\bf C}_{3}({\bf O})$
is not physically realistic.  It has the following four
diffficulties.

\vspace{12pt}

1.  The quadratic Jordan algebra action on
$J^{\bf C}_{3}({\bf O})$
mixes the elements ${\bf O}^{\bf C}_{+}$ and
${\bf O}^{\bf C}_{-}$ (fermion particles and
antiparticles) with ${\bf O}^{\bf C}_{v}$
(spacetime).

\vspace{12pt}

2.  The elements $a^{\bf C}, b^{\bf C}, c^{\bf C}$ have
no physical interpretation.

\vspace{12pt}

3.  The space $J^{\bf C}_{3}({\bf O})$ is complex,
not real.

The 8(complex) dimensional spacetime
${\bf O}^{\bf C}_{v}$ must be converted into
an 8(real) dimensional spacetime prior to its
dimensional reduction (as described in \cite{SM1,SM2})
to a physically realistic 4(real) dimensions.
\vspace{12pt}

4.  The space $J^{\bf C}_{3}({\bf O})$ is not
compact.

To do calculations, compact spaces are needed
so that finite volumes can be calculated.

The requirement of compact spaces is also
consistent with the use of compact projective
ray spaces as state spaces in the formalism of
matrix operators on Hilbert space.

\vspace{12pt}

To get around the first two of these difficulties,
it would be nice to have as space state a subspace of
$J^{\bf C}_{3}({\bf O})$ like:

\[
\left(
\begin{array}{ccc}
0 & {\bf O}^{\bf C}_{+} & 0  \\
& & \\
{\bf O}^{{\bf C}{\dagger}}_{+} & 0
 &  {\bf O}^{\bf C}_{-} \\
& & \\
0 & {\bf O}^{{\bf C}{\dagger}}_{-} & 0
    \end{array}
\right)
\oplus
\left(
\begin{array}{ccc}
0 & 0 &   {\bf O}^{\bf C}_{v} \\
& & \\
0 & 0 &  0 \\
& & \\
{\bf O}^{{\bf C}{\dagger}}_{v} & 0 & 0
\end{array}
\right)
\]

To find such a subspace, consider the HJTS
of $Type \: V$: $M^{\bf C}_{1,2}({\bf O})$.

\subsection{HJTS of $Type \: V$:
$M^{\bf C}_{1,2}({\bf O})$}

The full conformal Lie algebra corresponding to
the HJTS of $Type \: V$ is the 3-graded Lie
algebra
$$E^{\bf C}_{6} =$$
$$= M^{\bf C}_{1,2}({\bf O}) \oplus
(Spin^{\bf C}(10) \otimes U^{\bf C}(1))
\oplus M^{\bf C}_{1,2}({\bf O})$$.

\vspace{12pt}

The translations and the special conformal
transformations are each represented by
the space $M^{\bf C}_{1,2}({\bf O})$ of the
HJTS of $Type \: V$.

\vspace{12pt}

Since $M^{\bf C}_{1,2}({\bf O})$ can be
represented as

\[
\left(
\begin{array}{ccc}
0 & {\bf O}^{\bf C}_{+} & 0  \\
& & \\
{\bf O}^{{\bf C}{\dagger}}_{+} & 0
 &  {\bf O}^{\bf C}_{-} \\
& & \\
0 & {\bf O}^{{\bf C}{\dagger}}_{-} & 0
    \end{array}
\right)
\]

It seems to be a good candidate for the
part of the space of states representing
the fermion particles and antiparticles.

\vspace{12pt}

In particular, the fermion particles and
antiparticles can be mixed, as they should;
spacetime is not mixed, as it should not be;
and there are no $a^{\bf C}, b^{\bf C}, c^{\bf C}$
that have no physical interpretation.

\vspace{12pt}

However, $M^{\bf C}_{1,2}({\bf O})$ is complex
and is not compact, and so suffers from the
third and fourth dificulties.

To get a real and compact space of states,
consider the Silov boundary of the bounded
homogeneous circled domain corresponding to
the Hermitian symmetric space of $Type \: V$.

\subsubsection{Fermion Space of States =
$({\bf{R}}P^{1} \times S^{7}) \times ({\bf{R}}P^{1}
\times S^{7})$}

The $Type \: V$ Hermitian symmetric space of
compact type is Rosenfeld's elliptic projective plane,
the irreducible K\"{a}hler manifold
$${\bf{(C \otimes O)}}P^2 =
{E_{6} \over {Spin(10) \times U(1)}}$$
with 32 real dimensions. \cite{HLG,ABS}

\vspace{12pt}

The bounded homogeneous circled domain corresponding
to the $Type \: V$ HJTS is an irreducible bounded
complex domain of 32 real dimensions.

\vspace{12pt}

Its 16(real) dimensional Silov boundary, denoted here
by $S_{+} \times S_{-}$, is two copies of
${\bf{R}}P^{1} \times S^{7}$  \cite{McC,HLG,HUA}, or
$$S_{+} \times S_{-} =$$
$$= ({\bf{R}}P^{1} \times S^{7}) \times ({\bf{R}}P^{1}
\times S^{7})$$

\vspace{12pt}

$S_{+}$ = $S_{-}$ = ${\bf{R}}P^{1} \times S^{7}$ is the
compact 8(real) dimensional space of states that
represents the 8 (first generation) fermion particles
($S_{+}$) and antiparticles ($S_{-}$).

\vspace{12pt}

If $\{1,i,j,k,e,ie,je,ke\}$ is a basis (with real scalars)
for ${\bf{R}}P^{1} \times S^{7}$, where the real axis $\{1\}$
spans ${\bf{R}}P^{1}$ and $\{i,j,k,e,ie,je,ke\}$ spans
$S^{7}$, then

\vspace{12pt}

Since ${\bf{R}}P^{1}$ and $S^{7}$ are both parallelizable,
so is the whole space ${\bf{R}}P^{1} \times S^{7}$; and

Since ${\bf{R}}P^{1}$ is double covered by $S^{1}$, so that
if ${\bf{R}}P^{1}$ is parameterized by $U(1)$ then
$S^{1}$ is parameterized by $U(1) \times U(1)$:

Then, the representation of the (first generation) particles
by ${\bf{R}}P^{1}$ and $S^{7}$ as state space is

\[
\begin{array}{|c|c|c|} \hline
{\bf{R}}P^{1} \times S^{7}  & Fermion \: Particle & Phase\\
basis \: element & & \\ \hline
1 & e-neutrino & U(1)  \\ \hline
i & red \: up \: quark & U(1) \times U(1) \\ \hline
j & green \: up \: quark & U(1) \times U(1) \\ \hline
k & blue \: up \: quark & U(1) \times U(1) \\ \hline
e & electron & U(1) \times U(1) \\ \hline
ie & red \: up \: quark & U(1) \times U(1) \\ \hline
je & green \: up \: quark & U(1) \times U(1) \\ \hline
ke & blue \: up \: quark & U(1) \times U(1) \\ \hline
\end{array}
\]

The neutrino only has one $U(1)$ phase, and so should be
a Weyl fermion existing in only one helicity state.

The electron and quarks have $U(1) \times U(1)$ phases,
and so should be Dirac fermions existing in both left-
and right-handed helicity states.

The (first generation) fermion antiparticles are represented
similarly, except as mirror images of the particles.

\vspace{12pt}

The second and third generations of fermion particles and
antiparticles are formed by the spacetime dimensional
reduction mechanism as discussed in \cite{SM1,SM2}.

\vspace{12pt}

What about the spacetime part of the space of states?

The fermion part of the space of states was found
within the translation (or, equivalently, the special
conformal) part of the conformal Lie algebra of the
HJTS of $Type \: V$.

\vspace{12pt}

That leaves the Lorentz part (and the $U(1)$) of the
conformal Lie algebra of the HJTS of $Type \: V$.

The Lorentz part of the conformal Lie algebra of the
HJTS of $Type \: V$  turns out to be the conformal
Lie algebra of an HJTS of $Type \: IV_{7}$.

\vspace{12pt}

To find the spacetime part of the state space,
look at the translation (or, equivalently, the special
conformal) part of the conformal Lie algebra of the
HJTS of $Type \: IV_{7}$.

\vspace{12pt}

The Lorentz part of the conformal Lie algebra of the
HJTS of $Type \: IV_{7}$ will then be available to
construct the Lie algebra of the physical gauge
group, which can be shown to act in a physically
on the fermion part of the space of states just defined.

\vspace{12pt}

\subsection{HJTS of $Type \: IV_{7}$:
$J(\Gamma^{\bf C}(7))$}

The exceptional HJTS of $Type \: IV_{7}$ is the
Jordan algebra $J(\Gamma^{\bf C}(7))$  generated by
the Dirac gamma matrices $\Gamma^{\bf C}(7)$ in
7 dimensions with complex coefficients and the
Jordan triple product
\eq
(abc)= a\cdot (\bar{b} \cdot c) + c \cdot (\bar{b} \cdot a)
-(a \cdot c) \cdot \bar{b} \label{eq:HJTP}
\en
where the bar $\bar{\;}$ denotes complex
conjugation. \cite{MG2}

\vspace{12pt}

Since Clifford algebras are universal associative
algebfras, and since the Jordan algebra
$J(\Gamma^{\bf C}(7))$ is associative,
$J(\Gamma^{\bf C}(7))$ is contained in the even
subalgebra of the Clifford algebra $Clf({\bf C}^{8})$,
whose spin group is $Spin^{\bf C}(8)$ \cite{IP,NJ}.

\vspace{12pt}

The full conformal Lie algebra corresponding to
the HJTS of $Type \: IV_{7}$ is the 3-graded Lie
algebra
$$Spin^{\bf C}(10) =$$
$$= J(\Gamma^{\bf C}(7)) \oplus
(Spin^{\bf C}(8) \otimes U^{\bf C}(1))
\oplus J(\Gamma^{\bf C}(7))$$.

\vspace{12pt}

\subsubsection{Spacetime Space of States =
${\bf{R}}P^{1} \times S^{7}$}

The spacetime part of the space of states
should come from the translations of the
conformal Lie algebra of the HJTS of
$Type \: IV_{7}$.

\vspace{12pt}

The translations and the special conformal
transformations are each represented by
the space $J(\Gamma^{\bf C}(7))$ of the
HJTS of $Type \: IV_{7}$.

\vspace{12pt}

The $Type \: IV_{7}$ Hermitian symmetric space of
compact type is the irreducible K\"{a}hler manifold
$$Spin(10) \over {Spin(8) \times U(1)}$$
with 16 real dimensions \cite{HLG,ABS}.

\vspace{12pt}

The bounded homogeneous circled domain corresponding
to the $Type \: IV_{7}$ HJTS is an irreducible bounded
complex domain of 16 real dimensions.

\vspace{12pt}

Its 8(real) dimensional Silov boundary,
denoted here by $V_{8}$, is
${\bf{R}}P^{1} \times S^{7}$  \cite{McC,HLG,HUA}.

\vspace{12pt}

The 8(real) dimensional spaces $V_{8}$ = $S_{+}$ =
$S_{-}$ = ${\bf{R}}P^{1} \times S^{7}$ are all
isomorphic, as is required by the triality
property of $Spin(8)$ and $J^{\bf C}_{3}({\bf O})$.

\vspace{12pt}

\subsubsection{Gauge Group = $Spin(8)$}

The physical gauge group should be defined by the
Lorentz Lie algebra of the conformal Lie algebra
of the HJTS of $Type \: IV_{7}$, which is
$Spin^{\bf C}(8)$.

\vspace{12pt}

The complex  structure of the Lie algebra
$Spin^{\bf C}(8)$ gives the infinitesimal generators
of the $Spin(8)$ gauge group a complex $U(1)$ phase,
as is physically realistic for gauge boson propagators.

\subsubsection{Effect of Dimensional Reduction
on $Spin(8)$ Gauge Group}

As is discussed in \cite{SM1}, the effect of the
dimensional reduction of spacetime from
8(real) dimensions to 4(real) dimensions on the
gauge group $Spin(8)$ is to reorganize the 28
generators of $Spin(8)$ according to Weyl group
symmetry into the 10 generators of $Spin(5)$,
the 8 generators of $SU(3)$, the 6 generators
of $Spin(4)$, and the 4 generators of $U(1)^{4}$.

\vspace{12pt}

The $Spin(5)$, which is isomorphic to $Sp(2)$,
which is in some of the literature denoted by
$Sp(4)$,then gives gravity by the
MacDowell-Mansouri mechanism \cite{MM};

\vspace{12pt}

The $SU(3)$ gives the color force;

\vspace{12pt}

The $Spin(4)$ = $SU(2) \times SU(2)$  gives the
SU(2) weak force and, by integrating out the
other SU(2) over the 4(real) dimensions that
are eliminated by reduction of spacetime from
8(real) dimensions to 4(real) dimensions, gives
a Higgs scalar field for the Higgs mechanism; and

\vspace{12pt}

The $U(1)^{4}$ gives the 4 covariant components
of the electromagnetic photon.

\vspace{12pt}

Details of the above are in \cite{SM1}.

The details include such things as description
of the nonstandard relationship between the
weak force and electromagnetism, calculation
of the Weinberg angle, etc.

\vspace{12pt}

\section{Calculation of Observable Quantities}

The method for calculating force strengths is
based on the relationships of the four gauge groups
$Spin(5)$, $SU(3)$, $Spin(4)$, and $U(1)^{4}$
to the fundamental compact fermion and spacetime
state space manifolds.

Details are in \cite{SM1}.  A sketch is given here.

\vspace{12pt}

\subsection{Basis for Calculation}

The calculated strength of a force is taken to be
proportional to the product of four factors:
$$\left(1 \over \mu^2 \right) \left( Vol(M) \right)
\left( {Vol(Q)} \over {{Vol(D)}^{1 \over m}} \right)$$
where

\vspace{12pt}

$\mu$ is a symmetry breaking mass scale factor that
is

the Planck mass for gravity,

the weak symmetry
breaking scale for the weak force, and

$1$ for the color force and electromagnetism,
in which symmetry breaking does not occur.

\vspace{12pt}

$Vol(M)$ is the volume of the irreducible
$m$(real)-dimensional symmetric space on which
the gauge group acts naturally as a component of
4(real) dimensional spacetime $M^{4 \over m}$.

The $M$ manifolds for the gauge groups of
the four forces are:

\[
\begin{array}{|c|c|c|c|}
\hline
Gauge \: Group & Symmetric \: Space & m
& M  \\
\hline
& & \\
Spin(5) & Spin(5) \over Spin(4)  & 4 & S^4\\
& & \\
SU(3) & SU(3) \over {SU(2) \times U(1)}
& 4  & {\bf C}P^2 \\
& & \\
SU(2) & SU(2) \over U(1)  & 2 & S^2 \times S^2 \\
& & \\
U(1) & U(1)  & 1 & S^1 \times S^1 \times S^1
\times S^1 \\
& & \\
\hline
\end{array}
\]

\vspace{12pt}

$Vol(Q)$ is the volume of that part of the full compact
fermion state space manifold ${\bf R}P^1 \times S^7$
on which a gauge group acts naturally through its
charged (color or electromagnetic charge) gauge bosons.

\vspace{12pt}

For the forces with charged gauge bosons,

$Spin(5)$ gravity (prior to action of the
Macdowell-Mansouri mechanism),

$SU(3)$ color
force, and

$SU(2)$ weak force,

$Q$ is the Silov boundary of

the bounded complex homogeneous
domain $D$ that corresponds to

the Hermitian symmetric
space on which the gauge group
acts naturally as a local isotropy (gauge) group.

\vspace{12pt}

For $U(1)$ electromagnetism, whose photon carries
no charge, the factors $Vol(Q)$ and $Vol(D)$ do
not apply and are set equal to $1$.

\vspace{12pt}

The volumes $Vol(M)$, $Vol(Q)$, and $Vol(D)$ are
calculated with $M, Q, D$ normalized to unit radius.

The factor $1 \over {{Vol(D)}^{1 \over m}}$ is a
normalization factor to be used if the dimension of
$Q$ is different from the dimension $m$, in order to
normalize the radius of $Q$ to be consistent with the
unit radius of $M$.

\vspace{12pt}

The $Q$ and $D$ manifolds for the gauge groups of
the four forces are:

\[
\begin{array}{|c|c|c|c|c|}
\hline
Gauge & Hermitian & Type & m
& Q  \\
Group & Symmetric & of & & \\
& Space & D & & \\
\hline
& & & & \\
Spin(5) & Spin(7) \over {Spin(5) \times U(1)}
& IV_{5} &4 & {\bf R}P^1 \times S^4 \\
& & & & \\
SU(3) & SU(4) \over {SU(3) \times U(1)}
& B^6 \: (ball) &4 & S^5 \\
& & & & \\
SU(2) & SU(3) \over {SU(2) \times U(1)}
& IV_{3} & 2 & {\bf R}P^1 \times S^2 \\
& & & & \\
U(1) & -  & - & 1  & - \\
& & & & \\
\hline
\end{array}
\]

\vspace{12pt}

\subsection{Results of Calculations}

The relative strengths of the four forces
can be calculated from the formula
$$\left(1 \over \mu^2 \right) \left( Vol(M) \right)
\left( {Vol(Q)} \over {{Vol(D)}^{1 \over m}} \right)$$

\vspace{12pt}

The $1 \over \mu^2$ factor is only applicable to the
weak force and gravity.

For the weak force,
$${1 \over \mu^2} = {1 \over {m_{W+}^{2}
+m_{W-}^{2} + m_{Z}^{2}}}$$

For gravity,
$${1 \over \mu^2} = {1 \over {m_{Planck}^{2}}}$$

\vspace{12pt}

The geometric force strengths, that is, everything
but the symmetry breaking mass scale factors, are
$$\left( Vol(M) \right)
\left( {Vol(Q)} \over {{Vol(D)}^{1 \over m}} \right)$$
They are normalized by dividing them
by the largest one, the one for gravity.

\vspace{12pt}

The geometric volumes needed for the calculations,
mostly taken from Hua \cite{HUA}, are

\[
\begin{array}{||c||c|c||c|c||c|c||}
\hline
Force & M & Vol(M) & Q & Vol(Q) & D & Vol(D)  \\
\hline
& & & & & & \\
gravity & S^4 & 8\pi^{2}/3
& {\bf R}P^1 \times S^4  & 8\pi^{3}/3
& IV_{5} & \pi^{5}/2^{4} 5! \\
\hline
& & & & & &\\
color & {\bf C}P^2 & 8\pi^{2}/3
& S^5 & 4\pi^{3}
& B^6 \: (ball) & \pi^{3}/6 \\
\hline
& & & & & & \\
weak & {S^2} \times {S^2} & 2 \times {4 \pi}
& {\bf R}P^1 \times S^2 & 4 \pi^2
& IV_{3} & \pi^{3} / 24 \\
\hline
& & & & & & \\
e-mag  & T^4  & 4 \times {2\pi}
& -  & -
& -  & - \\
\hline
\end{array}
\]

Using these numbers, the results of the
calculations are the relative force strengths
at the characteristic energy level of the
generalized Bohr radius of each force:

\[
\begin{array}{|c|c|c|c|c|}
\hline
Gauge \: Group & Force & Characteristic
& Geometric & Total \\
& & Energy & Force & Force \\
& & & Strength & Strength \\
\hline
& & & & \\
Spin(5) & gravity & \approx 10^{19} GeV
& 1 & G_{G}m_{proton}^{2} \\
& & & & \approx 5 \times 10^{-39} \\
\hline
& & & & \\
SU(3) & color & \approx 245 MeV & 0.6286
& 0.6286 \\
\hline
& & & & \\
SU(2) & weak & \approx 100 GeV & 0.2535
& G_{W}m_{proton}^{2} \approx  \\
& & & & \approx 1.02 \times 10^{-5} \\
\hline
& & & & \\
U(1) & e-mag  & \approx 4 KeV
& 1/137.03608  & 1/137.03608 \\
\hline
\end{array}
\]

The force strengths are given at the characteristic
energy levels of their forces, because the force
strengths run with changing energy levels.

The effect is particularly pronounced with the color
force.

In \cite{SM1} the color force strength was calculated
at various energies according to renormalization group
equations, with the following results:

\[
\begin{array}{|c|c|}
\hline
Energy \: Level & Color \: Force \: Strength \\
\hline
&  \\
245 MeV & 0.6286 \\
& \\
5.3 GeV & 0.166 \\
& \\
34 GeV & 0.121  \\
& \\
91 GeV & 0.106 \\
& \\
\hline
\end{array}
\]

\vspace{1cm}
\normalsize
\footnoterule
\noindent
{\footnotesize {\bf Historical Note:}  As far as I know
the first attempt to calculate $\alpha_{E}$ by using
ratios of volumes of structures related to conformal
groups and bounded homogeneous complex domains was
by Armand Wyler (A. Wyler, {\it Arch. Ration. Mech. Anal.}
{\bf 31}, 35 (1968), A. Wyler, {\it Acad. Sci. Paris,
Comptes Rendus} {\bf 269A}, 743 (1969); A. Wyler,
{\it Acad. Sci. Paris, Comptes Rendus} {\bf 272A},
186 (1971):  see also {\it Physics Today} (August 1971)
17; {\it Physics Today} (November 1971) 9; and
R. Gilmore, {\it Phys. Rev. Lett.} {\bf 28}, 462 (1972)).}

\end{document}